\documentclass[10pt,a4paper]{article}

\usepackage[latin1]{inputenc}
\usepackage{times}
\usepackage{graphicx}
\usepackage{amsmath}
\usepackage{amsfonts}
\usepackage[latin1]{inputenc}
\usepackage[T1]{fontenc}
\usepackage{verbatim}
\usepackage{latexsym}
\usepackage{natbib}

\usepackage{hyperref}
\usepackage{url}

\hypersetup{backref,colorlinks=true,citecolor=blue}

\title{Determination of the most pertinent EUV proxy for use in thermosphere modeling}

\author{T. Dudok de Wit$^1$ and S. Bruinsma$^2$ \\
\small $^1 $ Laboratoire de Physique et Chimie de l'Environnement et de l'Espace,\\
\small UMR 6115 CNRS and University of Orl\'eans, 3A avenue de la Recherche Scientifique, 45071 Orl\'eans, France\\
\small $^2 $ DCT/SI/GS, Centre National d'Etudes Spatiales, 18 Avenue E. Belin, 31401 Toulouse, France}

\date{\normalsize \textit{This work has been published in Geophysical Research Letters 38 (2011) L19102, doi:10.1029/2011GL049028}}

\begin{document}
\maketitle

\sloppy

\begin{abstract}
Two major issues in the specification of the thermospheric density are the definition of proper solar inputs and the empirical modeling of thermosphere response to solar and to geomagnetic forcings. This specification is crucial for the tracking of low Earth orbiting satellites.

Here we address both issues by using 14 years of daily density measurements made by the Stella satellite at 813 km altitude and by carrying out a multiscale statistical analysis of various solar inputs. First, we find that the spectrally integrated solar emission between 26-34 nm offers the best overall performance in the density reconstruction. Second, we introduce linear parametric transfer function models to describe the dynamic response of the density to the solar and geomagnetic forcings. These transfer function models lead to a major error reduction and in addition open new perspectives in the physical interpretation of the thermospheric dynamics.
\end{abstract}


\section{Introduction}

The density and composition of Earth's thermosphere is mostly sensitive to variations of the solar irradiance in the Extreme UltraViolet (EUV, 10-121 nm) spectral range. EUV radiation heats the upper atmosphere, and an intensifying flux causes the density at a given altitude to increase. Such changes in the atmospheric density mainly affect objects in low Earth orbit, where the drag force becomes the second-largest (but secular) perturbation. 

Errors in upper atmosphere density models are one of the main reasons for the uncertainty in the knowledge of spacecraft and debris location, particularly so when tracking data are not available. A major source of error is the definition of the solar EUV forcing which, by lack of continuous observations until 2002, is customarily replaced by EUV proxies \citep{tobiska08,bowman08,lean09}, such as the $F_{10.7}$ index (the solar radio flux at 10.7 cm) and nowadays more frequently the $Mg II$ index (the core-to-wing ratio of the Mg II K-line at 280 nm). A second source of error is the static nature of the models, which precludes preconditioning due to lack of memory. 

 Here we show how a new approach allows both sources of error for thermospheric density nowcast to be reduced. First, we determine which single solar input is most appropriate by multiscale statistical analysis. Second, we introduce an empirical convolutive model that incorporates memory effects and allows both solar and geomagnetic forcings to be described simultaneously. This methodology can easily be extended to more than one solar input. 

The densities are inferred from precise orbit determination of the French geodetic Stella satellite, which is in a 96 $\deg$ inclination and near-circular orbit at approximately 813 km altitude. Stella is a suitable spacecraft for this kind of analysis because of its spherical shape (no attitude-related errors), the perfect knowledge of the satellite characteristics (mass, surface, reflectivity), and the very accurate laser tracking by the International Laser Ranging Service \citep{pearlman02}.


\section{The data}

In this study, we use 14 years of mean densities (from January 7, 1997 through July 31, 2010) derived from orbit perturbation analysis \citep{jacchia62} from Stella. The density is averaged over intervals of 24 hours. This data set has the advantage of being homogenous, with no averaging over various satellites. The first Drag Temperature Model (DTM) \citep{barlier78} is based upon such measurements, which essentially tie the observed decay of the semi-major axis to a mean density, which is estimated here with a relative uncertainty of 5\%. Seasonal variations, which are important at the altitude of Stella, are removed by windowed Fourier analysis. 

 The substitutes of the solar EUV flux we consider here are: 1) the $F_{10.7}$ index from Penticton Observatory, Canada; 2) the $MgII$ index from the LASP composite; 3) the integrated flux between 26-34 nm from the SEM radiometer onboard SoHO \citep{judge98}; 4) the $s10.7$ index, which has been built for orbitography purposes, using SEM data \citep{tobiska08}; 5) $Lya$, the intensity of the bright Lyman-$\alpha$ line (LASP composite); and 5) XUV, the baseline of the daily soft X-ray flux in the 0.1-0.8 nm band (from GOES). Data gaps in SEM are interpolated using a multivariate technique \citep{ddw11b}. Geomagnetic activity is represented by the planetary geomagnetic index $Ap$. Here, however, the focus is on testing solar inputs and not (yet) on optimizing the geomagnetic forcing.


\section{Determination of the best solar inputs}

Different time scales of the density are also associated with different physical mechanisms: fast variations are caused by geomagnetic activity and by solar rotation modulation of the EUV flux whereas longer time scales are associated with the lifetime of active regions and solar cycle. For that reason, we first decompose all quantities into a slowly-varying (DC) and a fluctuating (AC) component: $x(t) = x_{\rm DC}(t) + x_{\rm AC}(t)$.

The DC component is traditionally computed by running the data through a smoothing filter. An 81-day cutoff time is used in thermosphere models such as JB2008 \citep{bowman08b}, DTM2000 \citep{bruinsma03b} and MSIS \citep{picone02}. This smoothing, however, incorporates part of the fast variations in the DC component. This can be detrimental during geomagnetic storms, when sudden density bursts may cause the DC value to increase, see Fig.~1. We recommend instead the baseline or lower envelope, which is known to provide a better description of the slowly varying component in radio observations \citep{schmahl98}. We extract the baseline by taking the minimum value in a sliding 21-day window and subsequently smoothing that time series with a Gaussian filter that has a 21-day full width at half maximum. The major asset of the baseline is its resilience to peaks associated with geomagnetic storms, whose signature does not have to be removed manually.

\begin{figure}[!ht]
\noindent\includegraphics[width=0.75\textwidth]{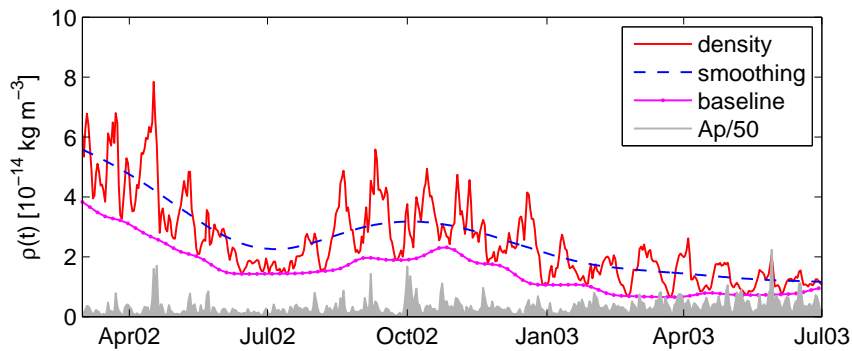}
\caption{Excerpt of the thermospheric density, showing the difference between the baseline and a Gaussian smoothing over 81 days.}
\label{fig_stella_1}
\end{figure}

The DC component of the density is found to be both in phase with and proportional to the solar forcing. Figure~2 shows a scatter plot of the DC component of the density $\rho$ versus that of solar inputs. To quantify the correlation between the two, we shall henceforth use two complementary descriptors: 1) Spearman's rank correlation $r$ is a direct measure of correlation \citep{wilks11}; we prefer it to the customary Pearson coefficient since it is invariant to nonlinear rescalings; 2) RMS is the classical Root Mean Square error, normalized to the standard deviation $\sigma_{\rho}$ of the density $\rho$, i.e. $\textrm{RMS} = \frac{1}{\sigma_{\rho}} \langle \left( \rho - \hat{\rho} \right)^2 \rangle^{1/2}$, where $\hat{\rho}$ is the modeled density; here the density is linear function of each solar proxy $x$, i.e. $\hat{\rho} = \alpha + \beta x$. Cross-validation is done by first estimating the model coefficients and then the RMS from independent subsets of the data. A RMS of 100\% means that none of the observed variability can be described by the model. Note also that a low RMS does not necessarily imply a high correlation, and vice-versa. Both are therefore needed to assess a solar input.

Based on the combined score of the correlation coefficient and RMS, we find from Fig.~2 that inputs such as the XUV flux and the sunspot number (not shown) can be readily excluded. We have selected $F_{10.7}$, $MgII$ and SEM for further analysis because they have the best scores but also because these quantities are guaranteed to remain available in the next decade and are best adapted for operational use. The scatter plots suggest that the DC component of the density $\rho$ can be relatively well modeled using a weakly non-linear function. For that reason, we define three new proxies labelled as $F$ (from $F_{10.7}$), $M$ (from $MgII$) and $S$ (from SEM). Each one is obtained by fitting the density with a cubic polynomial $\hat{\rho} = \alpha + \beta x + \gamma x^3$. Adding a quadratic or higher order terms does not reduce the RMS significantly with respect to its uncertainty. According to the RMS criterion, the best candidate for the DC component of the density is $S$, followed by $M$ and $F$. We find that the $S$ proxy also properly reproduces the density drop observed between the 1995-1996 and 2009-2010 solar minima, which supports the low EUV flux as being the primary cause of the low densities observed at the end of solar cycle 23 \citep{solomon10,solomon11}.

\begin{figure*}[!hb]
\noindent\includegraphics[width=\textwidth]{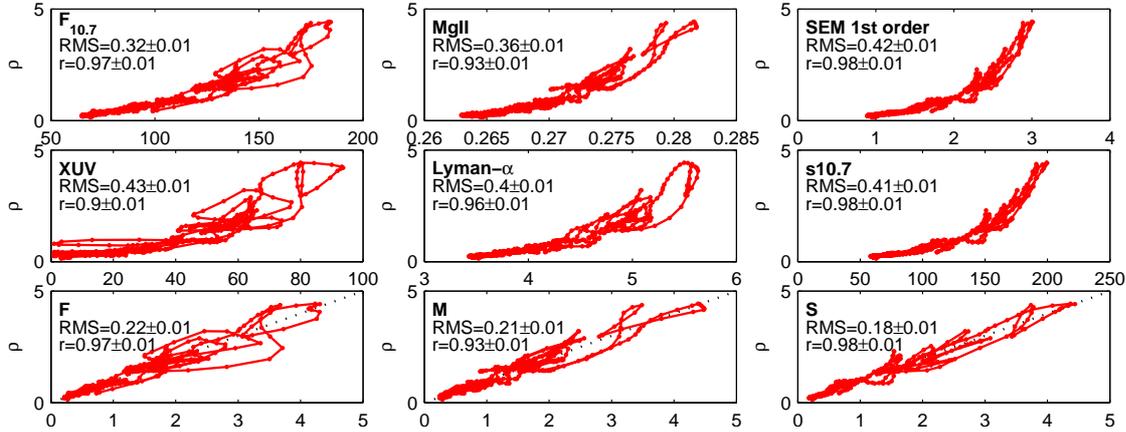}
\caption{Scatter plot of the DC component of the thermospheric density versus the DC component of various solar inputs. The RMS and Spearman's correlation coefficient $r$, together with their standard deviation, are indicated for each proxy. The standard deviation has been obtained by bootstrapping. In all plots, a linear model $\hat{\rho}_{\rm DC}= \alpha + \beta x_{\rm DC}$ is used for computing the RMS. Densities are in units of [$10^{-14}$ kg m$^{-3}$].)}
\label{fig_stella_2}
\end{figure*}

To compare the performance of individual solar inputs for short-term variations, we consider the multidimensional scaling technique used by \citet{ddw09}: all quantities are displayed on a 2D (so-called correspondence map) in such a way that their distance reflects their dissimilarity, which is expressed here by their pairwise RMS. The point of interest is the relative distance between quantities, not the axes. Such maps are widely used in statistics for their ability to provide a single global picture of the similarity between all the quantities.

Here, we compute the correspondence maps after using the \textit{\`a trous} wavelet transform to first decompose the AC component of each quantity into different time scales. By doing so, we investigate the similarities at different scales. Let us concentrate on three characteristic time scales that respectively correspond to half a solar rotation (i.e. center-to-limb effects), solar rotation and long-term effects, see Fig.~3. The latter is simply based on the DC component. At the shortest time scales of 1-2 days (not shown) the $Ap$ geomagnetic index is the quantity that is located closest to the density. At longer time scales, however, the shortest distances are systematically obtained with $S$, followed by $F$ or by $M$. From this, we conclude that the EUV flux in the 26-34 nm band, after a nonlinear rescaling, is the best overall solar proxy for the thermospheric density. The $F$ proxy, which is based on the widely used $F_{10.7}$ index, is a fallback option for time scales larger than a month, whereas the $M$ proxy is more suitable for short-term variations. This distinction highlights the importance of distinguishing different time scales. Other inputs, such as the intensity of the H Lyman-$\alpha$ line and the Magnetic Plage Strength Index systematically perform more poorly. 

 Interestingly, when correspondence maps show three aligned and closely-spaced quantities, then the quantity in the middle can be approximated by a linear combination of the two others. Figure 3 reveals that this is not the case with the density $\rho$, except for the largest scales. So, even though some improvement is possible by using more than one input (as in the JB2008 model), adding more inputs is unlikely to reduce the RMS further. Our representation thereby provides a strategy to determine the smallest combinations of inputs. 

\begin{figure*}[!ht]
\noindent\includegraphics[width=\textwidth]{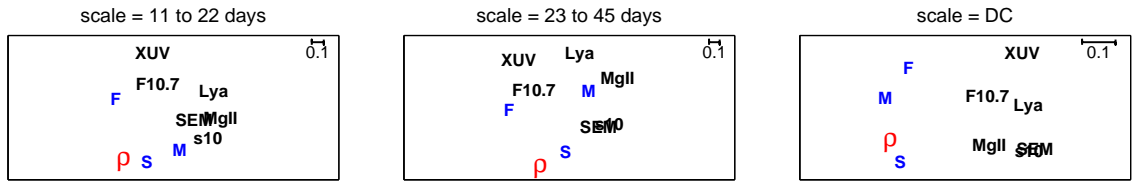}
\caption{Correspondence maps for three characteristic scales. The distance between each pair of points approximates their RMS error; axes have no immediate meaning. Characters correspond to the legend used in the text.}
\label{fig_stella_3}
\end{figure*}


\section{Model for the time-evolution}

Direct observations of the thermospheric density show that it does not respond instantaneously to external forcings but rather reacts with some delay. Two reasons for this are the inertia of the atmosphere and wavelength-dependent center-to-limb effects in the solar spectral irradiance. To incorporate this property in our reconstruction, we model the AC component using transfer functions by doing system identification (SI) \citep{ljung97}. We consider a particular class of discrete linear time-invariant models called \textit{Output Error} (OE), in which the modeled density $\hat{\rho}$ is a function of two inputs: $u_1$ (solar forcing) and $u_2$ (geomagnetic forcing), using the current date index $t$, and preceding days: 
\begin{equation} 
\hat{\rho}[t] = \frac{B_1(z^{-1})}{F_1(z^{-1})} u_1[t] + \frac{B_2(z^{-1})}{F_2(z^{-1})} u_2[t] 
\end{equation}
where $B_k(z^{-1}) = b_{k,1} + b_{k,2} z^{-1} + \cdots + b_{k,nb_k} z^{-nb_k+1}$, $F_k(z^{-1}) = 1 + f_{k,2} z^{-1} + \cdots + f_{k,nf_k} z^{-nf_k+1}$, and $z^{-1}$ is the delay operator of the $z$-transform, namely $z^{-1} u[t] = u[t-1]$. OE models are widely used to model linear systems with additive noise. We use information theoretic criteria to determine the optimum order of the model, and find typically $nf_1 = 2$, $nf_2 = 3$, $nb_1 = nb_2 = 3$, which means that 2 to 3 past values only are needed to describe the internal dynamics of the density (described by the $F(z^{-1})$ polynomials) and the response to the forcings (described by the $B(z^{-1})$ polynomials). 

OE models bring a major improvement over classical attempts to model the density. First, they provide a rigorous framework that contrasts with the (often subjective) selection criteria used to determine past and present combinations of solar inputs. Second, both solar and geomagnetic forcings can now be described simultaneously with a single model. This is a major improvement because, so far, all attempts to isolate the thermospheric response either to solar or to geomagnetic activity were severely constrained by the necessity to consider the very few intervals during which one of the forcings could be ignored, see for example \citep{sutton06}. So, by assuming that the response to the two inputs is linear and additive, we can now model the response of the density to any combination and any evolution of solar and geomagnetic activity levels. Incidentally, the OE model can give much deeper insight into the underlying physics, for example by providing access to the impulse or the step response of the density.

Figure 4 illustrates the good fit achieved by the OE model, and compares it with the DTM2000 and JB2008 models. To quantify the performance, we summarize in Table~\ref{performance} the RMS of the reconstructed density for various cases. We list three alternative models, but the OE model should be really compared to the static (i.e. memoryless) model only, because this is the only one that uses use exactly the same solar and geomagnetic inputs. The DTM2000 and JB2008 are only listed as examples of performance; the former uses the $F10.7$ and $Ap$ indices, and the latter four solar and two geomagnetic inputs. 

For the DC component of the density, the best performance is achieved with the $S$ proxy from SEM. Adding a second or a third solar proxy does not bring a major improvement, which supports our hypothesis that the RMS for long-term changes cannot be reduced further by using more solar forcing terms. Differences in the RMS, however, are also likely to be caused by the intrinsic long-term variability of the density, by possible instrumental drifts and by the modeling of the seasonal variation.

\begin{figure}[!hb]
\noindent\includegraphics[width=0.75\textwidth]{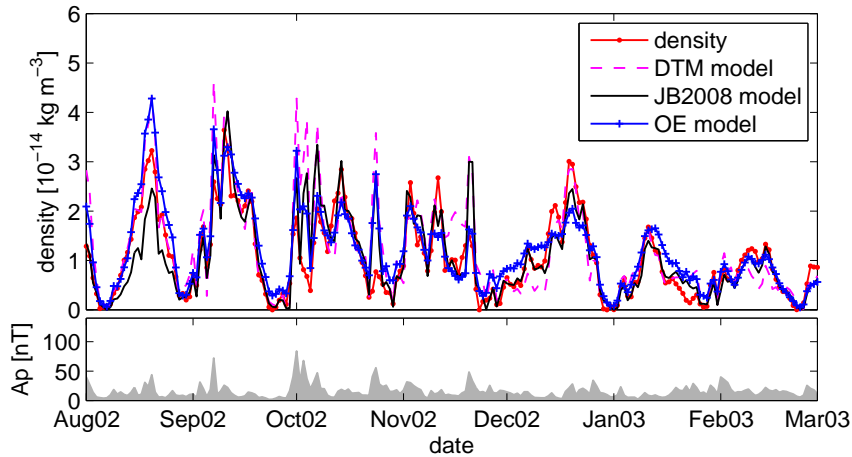}
\caption{Excerpt of the thermospheric density, also showing the fits from the DTM2000, the JB2008 and OE models.}
\label{fig_stella_4}
\end{figure}

The main improvement occurs in the AC component, which is also the one interest here. Not surprisingly, models that use several inputs (JB2008 or OE with 3 inputs) perform best. However, the OE model with one single input only ($S$) does almost as good and compared to the static case, the RMS is reduced by 20\%. This is the most important result of the table, as it highlights the good performance of our simple empirical model. This is again reflected in the global performance (DC \& AC), from which we conclude that $S$ is the best all-purpose solar input, way ahead of the $MgII$ and $F_{10.7}$ indices.

\begin{table}
\begin{center}
\begin{tabular}{c|c|ccc} \hline
model 		& solar input	 		& DC only 			& AC only 			& DC \& AC 		\\ \hline
OE 			&	 $F$ 				&	0.22			&	0.50 			&	0.33 	\\
OE 			&	 $M$				&	0.21			&	0.48 			&	0.32	\\
OE 			&	 $S$				&	\textbf{0.17} 			&	\textbf{0.42} 			&	\textbf{0.26} 	\\
static 		&	 $S$ 				&	\textbf{0.17}			&	0.52 			&	0.34	\\ \hline
DTM2000 	&	$F10.7$				&	0.18			&   0.50 			& 	0.36	\\
JB2008 		& 4 inputs 				&	\textbf{0.16} 			& 	\textbf{0.36} 			& 	 \textbf{0.25} 	\\
OE			& $F, M, S$				&	\textbf{0.16} 			& 	\textbf{0.40} 			&	 \textbf{0.26} 	\\ \hline
\end{tabular}
\end{center}
\caption{RMS of the reconstructed density, for different cases: the DC part only (left column), the AC part only (central column) and the full density (right column). The static case refers to the $S$ proxy with an instantaneous response of the density (no convolutive model). All other models use different (and generally more) inputs. Values strictly inferior to the median of each column are shown in boldface.}
\label{performance}
\end{table}


\section{Conclusions}

This study shows that major improvements can still be made in the methodology used for modeling of the thermospheric density response to external forcings. Here, we focused on the solar forcing, using 14 years of daily-mean density measurements made at 813 km altitude. 

We find that the EUV flux in the 26-34 nm band (as measured by SoHO/SEM) does systematically better than either the $F_{10.7}$ or the $MgII$ indices. The RMS on the reconstructed density is typically 20\% lower and the superiority of this proxy is observed at all time scales, including solar rotation and solar cycle. This EUV flux is presently measured by SDO/EVE and soon will be by GOES/EUVS, making it a good candidate for operational space weather applications. Our method also provides a visual strategy for selecting the best combinations of solar inputs. 

 For the first time transfer function models have been used to describe the dynamic response of the thermosphere to the solar and geomagnetic forcings, thereby casting this problem in the rigorous framework of SI. Using a linear \textit{output error} model, we find that the RMS can be reduced by 20\% compared to the equivalent static model. The SI framework brings numerous additional advantages. Primarily, it allows to describe the response to arbitrary temporal evolutions of the inputs without the need to isolate periods during which one the forcings only is active; that constraint has so far been a major impediment to the analysis of the thermospheric dynamics. Second, linear transfer function models allow to estimate the impulse response of the thermosphere, which provides deeper insight into its physical characteristics. This response differs from the one estimated using flares (e.g. \citep{sutton06}), because in the EUV flare spectra differ from daily averaged spectra. These aspects will be detailed in a forthcoming publication.


\subsubsection*{acknowledgments}
This study received funding from the European Community's Seventh Framework Programme (FP7-SPACE-2010-1) under the grant agreement nr. 261948 (ATMOP project, www.atmop.eu). SB is equally supported by GRGS. The following institutes are acknowledged for providing the data: Laboratory for Atmospheric and Space Physics (Boulder), National Geophysical Data Center (NOAA), the Space Sciences Center at the University of Southern California, and Space Environment Technologies. We also thank S. C. Solomon and B. R. Bowman for their review.



\end{document}